\documentclass[notitlepage,a4paper,aps,prd,onecolumn,superscriptaddress,nofootinbib,groupedaddress,longbibliography]{revtex4-1}
\usepackage{amsmath}
\usepackage{amsfonts}
\usepackage{amssymb}
\usepackage[utf8]{inputenc}
\usepackage[T1]{fontenc}
\usepackage[colorlinks=true]{hyperref}
\usepackage[dvipsnames]{xcolor}
\usepackage{graphicx}
\usepackage[normalem]{ulem}
\usepackage{enumerate}
\usepackage{tikz}

\usetikzlibrary{external}
\newtheorem{theorem}{Theorem}

\newtheorem{definition}[theorem]{Definition}

\begin{document}
\title{Relativistic kinetic gases as direct sources of gravity}

\author{Manuel Hohmann}
\email{manuel.hohmann@ut.ee}
\affiliation{Laboratory of Theoretical Physics, Institute of Physics, University of Tartu, W. Ostwaldi 1, 50411 Tartu, Estonia}

\author{Christian Pfeifer}
\email{christian.pfeifer@ut.ee}
\affiliation{Laboratory of Theoretical Physics, Institute of Physics, University of Tartu, W. Ostwaldi 1, 50411 Tartu, Estonia}

\author{Nicoleta Voicu}
\email{nico.voicu@unitbv.ro}
\affiliation{Faculty of Mathematics and Computer Science, Transilvania University, Iuliu Maniu Str. 50, 500091 Brasov, Romania}

\begin{abstract}
We propose a new model for the description of a gravitating multi particle system, viewed as a kinetic gas. The properties of the (colliding or non-colliding) particles are encoded into a so-called one-particle distribution function, which is a density on the space of allowed particle positions and velocities, i.e.\ on the tangent bundle of the spacetime manifold. We argue that an appropriate theory of gravity, describing the gravitational field generated by a kinetic gas, must also be modeled on the tangent bundle. The most natural mathematical framework for this task is Finsler spacetime geometry. Following this line of argumentation, we construct a coupling between the kinetic gas and a recently proposed Finsler geometric extension of general relativity. Additionally, we explicitly show how the general covariance of the action of the kinetic gas on the tangent bundle leads to a novel formulation of its energy-momentum conservation in terms of its energy-momentum distribution tensor.
\end{abstract}

\maketitle


\section{Introduction}\label{sec:intro}
An ensemble of a large number of $P$ individual interacting and gravitating point particles can be described on several levels of accuracy and detail:
\begin{itemize}
	\item The most precise description is to derive the trajectories for each individual particle from their mutual interactions directly, which in general is a too complex task to be accomplished in a reasonable amount of time and with a reasonable amount of computational effort.
	\item Instead of deriving the behaviour of all individual particles, one can consider them as a kinetic gas and describe their properties collectively, in terms of a so-called \emph{one-particle distribution function (1PDF)}. The 1PDF still contains the information about the velocity distribution of the different particles, but further information about each individual particle is averaged out.
	\item  Averaging the 1PDF over the velocities of the multiple particle system leads to its description as a fluid.
\end{itemize}

The gravitational field generated by the particles is usually obtained on the level of least accuracy in the above list. The energy-momentum tensor of a fluid is derived as second moment of the 1PDF with respect to the velocities of the particles, which then sources the Einstein equations. The dynamics of the fluid itself are given by the Euler equations, which  follow from the fact that the energy-momentum tensor must be covariantly constant.

Since the description of the $P$~particle system as a kinetic gas is more accurate and finer than its fluid description, the conjecture of this article is that the same is true for the description of the gravitational field generated by the $P$~particle system. One feature which for example would be taken into account in this way is the velocity distribution over the different gas particles. In the fluid description the velocity distribution is averaged out.

Technically the 1PDF is a function on the tangent bundle of spacetime, or in other words on the position and velocity space of the particles. It describes the number of particles in the gas and their trajectories. When averaged over the particle velocities, one can extract information about the pressure, density and energy-momentum of the resulting fluid description, see \cite{Israel,Ehlers2011} and \cite[Sec. 2.3]{Rezzolla2013}, or \cite{Sarbach:2013fya}. A geometric description of the kinetic gas and its dynamics on the tangent bundle has been investigated in \cite{Sarbach:2013uba} and was applied to the formation of accretion discs in~\cite{Rioseco:2016jwc}. Moreover, first post Newtonian corrections to the behaviour of a gravitating gas in a fixed gravitational behaviour have been studied \cite{Agon:2011mz,RamosCaro:2012rz}. What is missing in these approaches is the dynamical backreaction of the kinetic gas on the gravitational field, accordingly a dynamical equation which determines the gravitational field generated by the kinetic gas directly.

The aim of this article is to present how a kinetic gas, in terms of the 1PDF, can be directly coupled to gravity, and how the 1PDF is the source term of a gravitational field equation directly on the tangent bundle.

In order to couple the 1PDF to gravity without averaging, it is necessary to construct the dynamics of the gravitational field on the same space on which the 1PDF lives. Since gravity is encoded into the geometry of spacetime, this implies that we need a description of the geometry of spacetime on the tangent bundle. A natural mathematical framework for this task is Finsler geometry \cite{Finsler,Bucataru,Shen}.

The idea of using Finsler geometry as generalized geometry of spacetime has been considered in the literature for long \cite{Beem,Pfeifer:2011tk,Javaloyes:2018lex,Asanov,Bucataru,Goenner:2008rr,Hohmann:2018rpp}. Simultaneously, multiple attempts of finding Finsler generalisations of the Einstein equations have been made \cite{Asanov,Rutz,Li2014,Pfeifer,Hohmann:2018rpp,Minguzzi:2014fxa}. Besides mathematical difficulties in the precise consistent formulation of indefinite Lorentzian Finsler geometry, one most important open question, in all of the attempts of using Finsler geometry as physically viable extension of the geometry of spacetime, is how to couple the geometry to physical matter fields correctly. One particular obstacle is that a Finslerian geometry does not only depend on the points of spacetime, but also on its directions.

By studying the kinetic gas in the language of Finsler geometry, we turn this obstacle into an advantage. Extending the previous studies \cite{Hohmann:2015ywa,Hohmann:2015duq}, which considered the dynamics of a kinetic gas in Finsler geometric language in the context of cosmology, we formulate an action of the kinetic gas on the tangent bundle. Its general covariance, i.e.\ invariance under coordinate changes of the base manifold, yields a new formulation of energy-momentum conservation of a kinetic gas in terms of a quantity directly formulated on the tangent bundle - the energy-momentum distribution tensor - and it enables us to couple the gas to gravity in a simple way. The Finslerian extension of Einstein gravity which was proposed in~\cite{Pfeifer}, and further developed to mathematical rigour and consistency recently \cite{Hohmann:2018rpp}, provides the canonical gravitational field equation on the tangent bundle, which naturally can be sourced by the 1PDF. By constructing the coupling of a kinetic gas to the Finslerian geometry of spacetime explicitly, we demonstrate how physically viable matter leads to a Finslerian spacetime geometry dynamically and solve the problem of how to couple physical matter to a Finsler geometric theory of gravity. 

The future application of the gravitational field equation is expected to highly improve the understanding of systems that are described by gravitating fluids, such as the  universe as a whole in cosmology, ordinary and neutron stars, as well as accretion disks of black holes, by replacing the averaged gravitating fluid by the more accurate and finer notion of a kinetic gas.

We present our results as follows. In Section \ref{sec:FST} we recapitulate the main mathematical notions of Finsler spacetime geometry, which are necessary to describe the kinetic gas in terms of Finsler geometry in Section \ref{sec:KG}. The definition of the 1PDF is provided in Section \ref{ssec:1PDF}. Then we use the 1PDF to construct an action of the kinetic gas on the tangent bundle in Section \ref{ssec:gasact}, before we find the energy-momentum distribution tensor of the kinetic gas from the general covariance of the action in Section \ref{ssec:EMDT}. Finally, in Section \ref{sec:GasGravity} we derive the Finsler gravitational field equations coupled to the kinetic gas before we conclude in Section \ref{sec:conc}.

\section{The geometric setup}\label{sec:FST}
The stage on which we couple the kinetic gas to gravity is the tangent bundle $TM$ of a spacetime manifold $M$, which we assume to be of dimension $4$ here. However, the whole construction presented in this article can straightforwardly be generalized to any spacetime dimension. Any local coordinate chart $(U,x^a)$ on $M$ induces a local coordinate chart $(TU,(x^a,\dot x^a))$ on $TM$, where an element $\dot x\in TU$ is a vector $\dot x$ in some tangent space $T_xM, x\in U$, which in local coordinates can be expressed as $\dot x = \dot x^a \partial_a|_x$. This procedure associates the manifold induced coordinates $(x^a,\dot x^a)$ to $\dot x$. If there is no risk of confusion, we will sometimes omit the indices in the coordinate representation. The canonical coordinate basis of the tangent and cotangent spaces of $TM$ is denoted by $(\partial_a = \frac{\partial}{\partial x^a}, \dot \partial_a = \frac{\partial}{\partial \dot x^a})$ resp. $(dx^a, d\dot x^a)$.

\subsection{Finsler spacetimes}
To describe the kinetic gas and its coupling to gravity geometrically on the tangent bundle, we employ Finsler spacetime geometry and the Finsler spacetime geometric description of the gravitational dynamics  \cite{Hohmann:2018rpp,Pfeifer:2019wus,Pfeifer:2011tk,Pfeifer}.

The Finslerian geometry of spacetime is derived from the geometric clock of observers, resp. free point particle action for massive particles on trajectories $\gamma:\mathbb{R}\rightarrow M, \tau \mapsto \gamma(\tau)$, with tangent vectors denoted by $\dot \gamma =  \frac{\partial \gamma^a}{\partial \tau}\partial_a = \dot{\gamma}^a\partial_a$ and tangent bundle representation $(\gamma, \dot \gamma)$:
\begin{align}\label{eq:ppact}
	S[\gamma] = \int d\tau\ F(\gamma, \dot \gamma)\,,
\end{align}
where $F$ is a $1$-homogeneous function on $TM$ with respect to its second argument. Later, when we consider the relativistic kinetic gas in Section \ref{sec:KG}, we assume that the trajectories $\gamma(\tau)$ of the particles which constitute the gas extremize this length functional, i.e.\ they are the geodesics of \eqref{eq:ppact}.

To make the mathematical setup precise, we recall the definition of Finsler spacetimes with which we work here. Details about this definition can be found in \cite{Hohmann:2018rpp}, which was distilled from existing definitions in the literature~\cite{Javaloyes:2018lex,Lammerzahl:2012kw,Beem}.

\begin{definition}\label{def:FST}
	Let $\mathcal{A}$ be a conic subbundle of $TM$ such that $TM\setminus\mathcal{A}$ is of measure zero. A Finsler spacetime is a pair $(M,L)$, where $L:\mathcal{A}\rightarrow \mathbb{R}$ is a smooth function, called the Finsler-Lagrange function, which satisfies:
	\begin{itemize}
		\item $L$ is positively homogeneous of degree two with respect to $\dot x$: $L(x,\alpha \dot x) = \alpha^2 L(x,\dot x),\ \alpha \in \mathbb{R}^+$;
		\item on $\mathcal{A}$ the vertical Hessian
		\begin{align}\label{g_def}
		g^L_{ab}=\dfrac{1}{2}\dfrac{\partial ^{2}L}{\partial \dot{x}^{a}\partial \dot{x}^{b}}
		\end{align}
		of $L$ (called $L$-metric) is non-degenerate in every coordinate chart;
		\item there exists a connected component $\mathcal{T}$ of the preimage $L^{-1}((0,\infty))\subset TM$ on which $g^L$ exists, is smooth and has Lorentzian signature $(+,-,-,-)~$\footnote{It is possible to equivalently formulate this property with opposite sign of $L$ and metric $g^L$ of signature $(-,+,+,+)$. We fixed the signature and sign of $L$ here to simplify the discussion.}
		\item the Euler-Lagrange equations
		\begin{align}\label{eq:EL}
		\frac{d}{d \tau} \dot{\partial}_a L - \partial_a L = 0\,.
		\end{align}
		have a unique local solution for every initial condition $(x,\dot x)\in \mathcal{T}\cup \mathcal{N}$, where $\mathcal{N}$ is the kernel of $L$. At points of $\mathcal{N}\setminus \mathcal{A}$, i.e.\ where the $L$-metric degenerates or does not even exist, the solution must be constructed by continuous extension. This means that the geodesic equation coefficients admit a $\mathcal{C}^{1}$ extension at those points.
	\end{itemize}
\end{definition}
The $1$-homogeneous function $F$, which defines the point particle action \eqref{eq:ppact}, is derived from the Finsler Lagrange function as $F=\sqrt{|L|}$. For clarity, we list the different sets which appear in the definition and comment on their meaning:
\begin{itemize}
	\item $\mathcal{A}$: the subbundle where $L$ is smooth and $g^L$ is non-degenerate, with fiber $\mathcal{A}_{x} = \mathcal{A} \cap T_xM$, called the set of \emph{admissible vectors},
	\item $\mathcal{N}$: the subbundle where $L$ is zero, with fiber $\mathcal{N}_x = \mathcal{N} \cap T_xM$,
	\item $\mathcal{A}_0 = \mathcal{A}\setminus\mathcal{N}$: the subbundle where $L$ can be used for normalization, with fiber  $\mathcal{A}_{0x} = \mathcal{A}_0 \cap T_xM$,
	\item $\mathcal{T}$: a maximally connected conic subbundle where $L > 0$, the $L$-metric exists and has Lorentzian signature $(+,-,-,-)$, with fiber $\mathcal{T}_x = \mathcal{T} \cap T_xM$.
\end{itemize}
The connected component $\mathcal{T}$ is interpreted as the set of future directed timelike directions on spacetime. An important subset of the timelike directions is formed by the future pointing unit timelike vectors
\begin{align}
\mathcal{O}:=\{(x,\dot x)\in \mathcal{T}| L(x,\dot x) = 1\}\,,
\end{align}
which is called the \emph{observer space} \cite{Gielen:2012pn,Gielen:2012fz,Hohmann:2013fca,Hohmann:2014gpa,Hohmann:2015pva}. This set is itself a fibered manifold over $M$, with fibers $\mathcal{O}_x = \mathcal{O} \cap T_xM$.

Finsler spacetimes are a straightforward generalisation of pseudo-Riemannian spacetimes equipped with a metric \(g_{ab}\) of Lorentzian signature. The latter constitute a subclass of Finsler spacetimes $(M,L)$ with $L = g_{ab}(x)\dot x^a \dot x^b$, defined by the components of a pseudo-Riemannian metric.

\subsection{Geometry of Finsler spacetimes}\label{ssec:geom}
Finsler geometry is a long-standing straightforward generalisation of Riemannian geometry \cite{Shen,Bucataru}. Finsler spacetime geometry is, up to the precise definition of Finsler spacetimes, an equally straightforward generalisation of pseudo-Riemannian geometry. Here we recall the geometric notions we need to describe the kinetic gas consistently in the language of Finsler geometry and to couple it as source term to the Finsler gravitational field equation.

Starting from the Finsler Lagrange function $L$, we can define canonical tensor fields on $\mathcal{A}_0\subset TM$. The Hilbert form\footnote{Mathematically precise, the restriction of $\omega$ to the observer space $\mathcal{O}$ defines a contact structure on $\mathcal{O}$.} $\omega$, the Finsler metric tensor $g^L$ and the Cartan tensor $C$, using the notation $\dot x_a = g^L_{ab}(x,\dot x)\dot x^b$, are,
\begin{align}
\omega =  \dot{\partial}_aF dx^a= \frac{L}{|L|}\frac{\dot x_a}{\sqrt{|L|}} dx^a,\quad g^L = g^L_{ab} dx^a \otimes dx^b,\quad C = \frac{1}{2}\dot \partial_c g^L_{ab} dx^a \otimes dx^b \otimes dx^c\,.
\end{align}

The geodesic equations of \eqref{eq:ppact} in arc length parametrization are given by the Euler-Lagrange equations \eqref{eq:EL} and take the form
\begin{align}
	\ddot x^a + 2 G^a(x,\dot x) = 0\,.
\end{align}
They are defined in terms of the so-called spray coefficients $G^a$, which in turn determine the Cartan non-linear connection coefficients $G^a{}_b$
\begin{align}\label{def:Ga}
	G^a = \frac{1}{4} g^{Lab}\left(\dot x^m \partial_m \dot \partial_b L - \partial_b L \right),\quad  G^a{}_b :=\dot \partial_b G^a\,.
\end{align}

The connection coefficients give rise to a splitting of the tangent and cotangent spaces to $\mathcal{A}$, into so-called horizontal and vertical parts. Locally, they are spanned by the basis vector fields
\begin{align}
	T_{(x,\dot x)}\mathcal{A} = \textrm{span}\{\delta_a = \partial_a - G^b{}_a\dot \partial_b;\ \dot{\partial}_a\}_{a=0}^3\,,\quad
	T^*_{(x,\dot x)}\mathcal{A} = \textrm{span}\{dx^a;\ \delta\dot x^a = dx^a + G^a{}_b  d\dot x^b\}_{a=0}^3\,.
\end{align}
With help of this adapted basis, on $\mathcal{A}_0$ we can introduce the following vector field, which is dual to the Hilbert form,
\begin{align}
	\mathbf{r}=\frac{\dot x^a}{\sqrt{|L|}} \delta_a,\quad \mathbf{i}_\mathbf{r}\omega = \omega(\mathbf{r}) = 1\,.
\end{align}
The restriction of  $\mathbf{r}$ to the observer space $\mathcal{O}$ is called the Reeb vector field $\mathbf{r}|_\mathcal{O}$ associated to $\omega$. By Cartan's magic formula, which relates the interior product, the exterior derivative and the Lie derivative $\mathfrak{L}_X = \mathbf{i}_{X}d + d\mathbf{i}_{X}$, and by the explicit expansion of $d\omega$ in the adapted basis the following important relation is satisfied
\begin{align}
	\mathfrak{L}_{\mathbf{r}}\omega = \mathbf{i}_\mathbf{r}d\omega + d\mathbf{i}_\mathbf{r}\omega = \mathbf{i}_\mathbf{r}d\omega = 0 \,.
\end{align}

The curvature tensor $\mathfrak{R}$ of the Cartan non-linear connection and the Finsler Ricci scalar $R_0$ are defined as
\begin{align}
	\mathfrak{R} = \frac{1}{2}R^a{}_{bc} dx^b \wedge dx^c \otimes \dot \partial_a,\quad R^a{}_{bc} \dot \partial_a  = (\delta_b G^a{}_c - \delta_c G^a{}_b)\dot{\partial}_a = [\delta_b,\delta_c],\quad R_0 = \frac{1}{L}R^a{}_{ab}\dot x^b\,.
\end{align}

Last but not least, the canonical volume form on the set $\mathcal{A}_0$ of a Finsler spacetime is given by
\begin{align}
	\textrm{Vol}_0 = \frac{|\det g^L|}{L^2} d^4x \wedge d^4\dot x := \frac{|\det g^L|}{L^2} dx^0 \wedge ... \wedge dx^3 \wedge d\dot x^0 \wedge ... \wedge d\dot x^3\,.
\end{align}
Denote by $\mathbb{C} = \dot x^a \dot \partial_a$ the Liouville vector field. It gives rise to a $7$-form $\Sigma$, see also \cite{Hohmann:2018rpp},
\begin{align}\label{eq:vol67}
   \Sigma
   &= \mathbf{i}_{\mathbb{C}}\textrm{Vol}_0  = \frac{1}{3!} \omega \wedge d\omega \wedge d\omega \wedge d\omega  = \frac{|\det g^L|}{L^2}\mathbf{i}_{\mathbb{C}}(d^4x \wedge d^4\dot x)\\
   &= \frac{1}{3!}\frac{|\det g^L|}{L^2} \dot x^a\epsilon_{abcd} dx^0 \wedge ... \wedge dx^3 \wedge d\dot x^b \wedge d\dot x^c \wedge d\dot x^d\,,
\end{align}
and a $6$-form $\Omega$
\begin{align}
	\Omega = \mathbf{i}_\mathbf{r}\Sigma = \frac{1}{3!} d\omega \wedge d\omega \wedge d\omega\,,
\end{align}
which obey
\begin{align}\label{eq:osprop}
	d\Omega=0,\quad \Sigma = \omega\wedge\Omega\,.
\end{align}
These differential forms will play the role of canonical volume forms on different subsets of the observer space $\mathcal{O}$.

In the literature on Finsler geometry several canonical linear covariant derivatives on $TM$ are considered, which all reduce to the Levi-Civita connection in the case of Pseudo-Riemannian geometry, i.e.\,, Finsler geometry with a Finsler Lagrangian that is quadratic in the dependence on $\dot x$, i.e.\ $L=g_{ab}(x)\dot x^a \dot x^b$. We employ the so called Chern-Rund linear covariant derivative, which can be defined by its action in the adapted basis as
\begin{align}\label{eq:CRCD}
	\nabla_{\delta_a}\delta_b = \Gamma^c{}_{ab}\delta_c,\ \nabla_{\delta_a}\dot \partial_b = \Gamma^c{}_{ab}\dot \partial_c,\ \nabla_{\dot \partial_a}\delta_b = 0,\ \nabla_{\dot \partial_a}\dot \partial_b = 0\,,
\end{align}
with $\Gamma^c{}_{ab} = \frac{1}{2}g^{Lcq}(\delta_a g^L_{bq}+\delta_b g^L_{aq}-\delta_q g^L_{ab})$. It satisfies the important identities, see \cite[p.104]{Bucataru},
\begin{align}\label{eq:CRprop}
	\nabla_{\delta_a}\dot x^b = 0, \nabla_{\delta_a}L = 0 \textrm{ and } \nabla_{\delta_a}g^L{}_{bc} = 0\,,
\end{align}
which we prove for completeness in Appendix \ref{app:CRConn}. Furthermore there exists the so-called dynamical covariant derivative, which is uniquely defined in terms of the canonical Cartan non-linear connection alone, independently of the choice of linear connection; we denoted the dynamical covariant derivative by $\nabla$ without any index. Operationally, it can be understood as
\begin{align}\label{eq:dynCD}
 \nabla\delta_b := \dot x^a \nabla_{\delta_a}\delta_b = G^c{}_b\delta_c,\ \nabla\dot \partial_b = \dot x^a\nabla_{\delta_a}\dot \partial_b = G^c{}_{b}\dot \partial_c\,.
\end{align}
Last but not least, we recall the components of the Landsberg tensor, in terms of the connection coefficients of the Chern-Rund connection or equivalently with help of the dynamical covariant derivative of the Cartan tensor,
\begin{align}\label{eq:Lands}
	P^a{}_{bc} = \dot{\partial}_c G^a{}_b - \Gamma^a{}_{cb} = g^{Lad}\nabla C_{dbc}\,.
\end{align}
The trace $P_b=P^a{}_{ab}$ will be part of the geometry side of the gravitational field equation \eqref{eq:fgravgas}.

With the help of the geometric setup just introduced, we can now describe a relativistic kinetic gas.

\section{The kinetic gas in the language of Finsler geometry}\label{sec:KG}
Instead of describing a relativistic gas, i.e. a collection of relativistic particles, particle by particle, the kinetic gas theory employs a 1PDF to describe the gas particles collectively. This approach to the description of the gas is more accurate than its approximation as a fluid. Classical reference to the topic are the article by Israel \cite{Israel} and the lectures by Ehlers~\cite{Ehlers2011}. A modern review can be found in the monograph \cite{Rezzolla2013} or in the articles \cite{Sarbach:2013fya,Sarbach:2013uba}.

In terms of the 1PDF, kinetic gases are naturally described on the tangent bundle of a manifold, which is the same stage on which Finsler geometry naturally lives. In this section, we introduce the 1PDF and use the geometry of Finsler spacetimes to deduce the action of a kinetic gas on the observer space $\mathcal{O}$ over $M$. Moreover, we study the conservation laws which follow from the invariance of the action under arbitrary 1-parameter groups of spacetime induced coordinate changes on $TM$. The arising Noether currents point out the novel notion of an energy-momentum distribution tensor of the gas on the tangent bundle, via a Gotay-Marsden type procedure \cite{GotayMarsden,Voicu:EM}. We will relate this new notion of energy-momentum to the usual definition of the energy-momentum tensor of a kinetic gas on the spacetime manifold $M$.

The analysis of the properties of the tangent bundle action of a kinetic gas prepares its coupling to Finsler gravity in the next section.

\subsection{General description}\label{ssec:1PDF}
A kinetic gas is a collection of $P$ particles which propagate through spacetime on piecewise normalized geodesics~$\gamma(\tau)$. In the language of a Finsler spacetime, this means that the tangent vectors of the trajectories of particles are elements of the observer space, $\frac{d\gamma}{d\tau}=\dot{\gamma}(\tau)\in \mathcal{O}_{\gamma(\tau)} \subset \mathcal{T}_{\gamma(\tau)}$ and that the tangents of the lifted trajectories $c(\tau)=(\gamma(\tau), \dot{\gamma}(\tau))$ are given by the Reeb vector field $\dot c= \mathbf{r}|_c$. The latter statement is equivalent to saying that the particle trajectories extremize the point particle action integral
\begin{align}\label{eq:gasactpp}
S[\gamma] = m \int_{\tau_1}^{\tau_2} c^*\omega = m \int_{\tau_1}^{\tau_2} \frac{\dot \gamma_a}{\sqrt{|L(\gamma,\dot \gamma)|}} c^*(dx^a) = m\int_{\tau_1}^{\tau_2} \frac{\dot \gamma_a}{\sqrt{|L(\gamma,\dot \gamma)|}} \dot \gamma^a d\tau = m \int_{\tau_1}^{\tau_2} d\tau\ F(\gamma,\dot \gamma) = m t\,.
\end{align}
It is defined by a Finsler function $F$, which in turn is derived from a Finsler spacetime Lagrange function $L$, see Definition~\ref{def:FST}. The number $t = \tau_2 - \tau_1$ denotes the proper time passing along a particle trajectory between $\gamma(\tau_1)$ and $\gamma(\tau_2)$.

Instead of describing the motion of all particles individually, the kinetic gas theory employs the 1PDF,
\begin{align}
\phi: \mathcal{O} \rightarrow \mathbb{R};\quad  (x,\dot x)\mapsto \phi(x,\dot x)\,,
\end{align}
which expresses the number \(N[\sigma]\) of particle trajectories passing through an oriented, 6-dimensional hypersurface $\sigma\subset \mathcal{O}$ through the integral
\begin{align}\label{eq:N}
N[\sigma] = \int_{\sigma} \phi \Omega\,.
\end{align}
It vanishes for hypersurfaces for which the tangent vectors $\dot c$ are tangent to $\sigma$, i.e.\ $\dot c \in T_c \sigma$. The integral is non-vanishing for hypersurfaces that are transversal to the particle trajectories, i.e.\ for which $\dot c \notin T_c \sigma$. The canonical volume form $\Omega$ on $\sigma$ was defined in \eqref{eq:vol67}. Since in all practical physical situations, there will always be gas particles with a finite maximal velocity, we assume in what follows that for all $x\in M$ $\phi_x(\dot x) = \phi(x,\dot x)$ has compact support on the set of future pointing unit timelike directions $\mathcal{O}_x$.

An important feature of this integral is that its result is independent of the geometric field $L$ on $TM$\footnote{Note that the integrand in \eqref{eq:N} is $0$-homogeneous with respect to its $\dot x$ dependence, i.e. it depends only on the future pointing \emph{directions} on $M$. In \cite{Hohmann:2018rpp} we have shown that such integrals over compact domains on the unit tangent bundle $U=\{(x,\dot x)\in TM| |L(x,\dot x)| = 1\}$ are identical to integrals over compact domains on the so called projectivised tangent bundle. In particular, this holds for compact domains in $\mathcal{O}\subset U$. While $U$ and $\mathcal{O}$ are defined in terms of $L$, the projectivised tangent bundle is not. Hence, being able to map the integration domain $\sigma$ of the number counting integral to a subset $\sigma^+$ of the projectivised tangent bundle (that is independent of $L$) allows us to commute the variation with respect to $L$ with integration. Here we do not discuss these  mathematical details, which will be reviewed in detail an upcoming article on the mathematical foundations of field theory on Finsler spacetimes.}. It only depends on the trajectories of the particles and the hypersurface chosen. Therefore, its variation with respect to the Finsler Lagrangian vanishes and we find the equation
\begin{align}\label{eq:dLN=0}
\delta_L N[\sigma] = \int_{\sigma} \delta_L(\phi\ \Omega) = \int_{\sigma} ( \delta_L\phi \Omega + \phi \delta_L \Omega) = 0\,.
\end{align}
It defines a relation between the 1PDF $\phi$ and the Finsler Lagrangian $L$, which is the important feature that couples the gas to the geometry of spacetime.

\begin{figure}[h!]
	\begin{tikzpicture}[thick]
	\foreach \x in {-5,-4,-3,-2} \draw (\x,-0.5) .. controls +(0.25,0.5) and +(0.25,-0.5) ..  (\x,1.5);
	\filldraw[Green,fill=white] (-6,0) .. controls (-5,0.5) and (-2,0.5) .. (-1,0) .. controls (-0.5,0.5) and (0,1) .. (0,1.5) node [right] {$\sigma_0$} .. controls (-1,2) and (-4,2) .. (-5,1.5) .. controls (-5,1) and (-5.5,0.5) .. (-6,0);
	\foreach \n / \x / \y in {1/-4.5/1.0,2/-3.5/1.5,3/-2.5/1.0,4/-1.5/1.5} \draw (\x,\y) .. controls +(0.25,1) .. +(0,2) node [at start,right] {$c_{\n}(0)$};
	\foreach \x / \y in {-4.5/1.0,-3.5/1.5,-2.5/1.0,-1.5/1.5} \fill[Green] (\x,\y) circle [radius=1.5pt];
	\filldraw[Blue,fill=white] (-6,2) .. controls (-5,2.5) and (-2,2.5) .. (-1,2) .. controls (-0.5,2.5) and (0,3) .. (0,3.5) node [right] {$\sigma_t$} .. controls (-1,4) and (-4,4) .. (-5,3.5) .. controls (-5,3) and (-5.5,2.5) .. (-6,2);
	\foreach \n / \x / \y in {1/-4.5/3.0,2/-3.5/3.5,3/-2.5/3.0,4/-1.5/3.5} \draw (\x,\y) .. controls +(0.1,0.5) .. (\x,4.5) node [at start,right] {$c_{\n}(t)$};
	\foreach \x / \y in {-4.5/3.0,-3.5/3.5,-2.5/3.0,-1.5/3.5} \fill[Blue] (\x,\y) circle [radius=1.5pt];
	\draw[Green] (1,0) .. controls (2,0.5) and (5,0.5) .. (6,0) .. controls (6.5,0.5) and (7,1) .. (7,1.5) node [right] {$\sigma_0$} .. controls (6,2) and (3,2) .. (2,1.5) .. controls (2,1) and (1.5,0.5) .. (1,0);
	\foreach \pos in {(1,0),(6,0),(7,1.5),(2,1.5)} \draw \pos .. controls +(0.25,1) .. +(0,2);
	\filldraw[Blue,fill=white] (1,2) .. controls (2,2.5) and (5,2.5) .. (6,2) .. controls (6.5,2.5) and (7,3) .. (7,3.5) node [right] {$\sigma_t$} .. controls (6,4) and (3,4) .. (2,3.5) .. controls (2,3) and (1.5,2.5) .. (1,2);
	\draw (3.5,2.1) node {$V$};
	\end{tikzpicture}
	\caption{Construction of hypersurfaces by the flow of the Reeb vector field, and the corresponding swept out volume.}
	\label{fig:vol}
\end{figure}
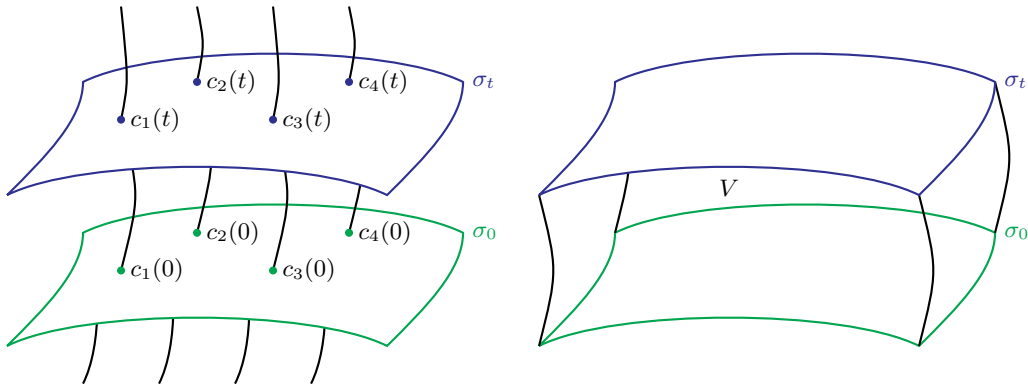

For a collisionless gas, the Liouville equation holds
\begin{align}\label{eq:lville}
	\mathbf{r} (\phi) =  \dot x^a \delta_a \phi = 0\,,
\end{align}
which can be seen by the following argument, see for example \cite{Sarbach:2013fya,Hohmann:2015ywa}. Choose a hypersurface $\sigma_0\subset \mathcal{O}$ as above. We obtain a family of hypersurfaces $\sigma_s$ by following the flow of the Reeb vector field $\mathbf{r}$ from each point in $\sigma_0$ for one and the same parameter $0<s< t$. This family of hypersurfaces spans a volume $V=\bigcup_{s\in (0, t)}\sigma_s$, see Figure \ref{fig:vol} for a sketch.
The difference between the number of particles on $\sigma_0$ and $\sigma_t$ is given by
\begin{align}\label{eq:Ns1s2}
	N[\sigma_t] - N[\sigma_0] = \int_V \mathbf{r}(\phi)\ \Sigma\,.
\end{align}
For a collisionless gas $N[\sigma_t] - N[\sigma_0] = 0$ and hence the Liouville equation  \eqref{eq:lville} follows. The proof of  \eqref{eq:Ns1s2} involves Stokes' Theorem and the properties \eqref{eq:osprop} of the forms $\Omega$ and $\Sigma$, details can be found in \cite{Sarbach:2013fya,Hohmann:2015ywa}.

With help of the 1PDF and the number counting integral, we can now construct the action of a kinetic gas on $TM$.

\subsection{The action of a kinetic gas}\label{ssec:gasact}
Let us consider a kinetic gas which consists of $P$ individual particles of equal mass $m$. Hence the action of a gas in a volume $V$ generated by the flow of the Reeb vector field from an initial hypersurface $\sigma_0$, which is pierced by all $P$ particle trajectories, to a final hypersurface $\sigma_t$, with flow parameter $0 \leq s \leq t$, $V=\bigcup_{0<s<t}\sigma_s \subset U$, see Figure \ref{fig:vol}, is given by
\begin{align}\label{eq:gasact1}
	S_{\text{gas}} = m P t = m P\int_{\tau_1}^{\tau_2} d\tau\ F(\gamma,\dot \gamma) = P \int_{\tau_1}^{\tau_2} c^*\omega = m \int_0^t P ds\,.
\end{align}

To express the action in terms of the 1PDF, we will express the particle number $P$ in \eqref{eq:gasact1} by the number counting integral $P=N[\sigma_s]$ to obtain
\begin{align}
	S_{\text{gas}} = m \int_0^t \left( \int_{\sigma_s} \phi \Omega \right) ds = m \int_V \phi \Omega \wedge \omega \,.
\end{align}
Applying the relation \eqref{eq:osprop}, which relates the different volume forms, we conclude that the tangent bundle action of a kinetic gas is
\begin{align}\label{eq:gasact}
	S_{\text{gas}} = m \int_V \phi \Sigma\,.
\end{align}
This action depends on the particle trajectories $(\gamma(\tau),\dot \gamma(\tau))$, which determine the 1PDF $\phi(x,\dot x)$ from the number integral \eqref{eq:N}, and on the geometry defining Finsler Lagrange function $L$. It thus is defined by the Lagrangian $7$-form
\begin{align}\label{eq:L7form}
 \lambda(x,\dot x, L(x,\dot x), \dot{\partial}\dot{\partial}L(x,\dot x)) = m\phi\Sigma\,.
\end{align}
Variation of the action with respect to the particle trajectories yields the Finsler geodesic equation for each of the particles, while variation with respect to the Finsler Lagrange function yields the source term for the gravitational dynamics.

We like to remark that the volume $V$ over which the action integral is taken is usually assumed to be composed out of compact domains $D\subset M$ and $V_x\subset \mathcal{O}_x$ for each $x \in D$, where $V_x$ is chosen such that it contains the support of $\phi_x$, i.e.\ $V = \bigcup_{x\in D}V_x$. With such choice of $V$, it is possible to split the action integral
\begin{align}\label{eq:actsplit}
S_{\text{gas}} = m \int_V \phi \Sigma = m \int_D \left(\int_{V_x} \phi_x(\dot x) \Sigma_x\right) d^4x = m \int_D \left(\int_{\mathcal{O}_x} \phi_x(\dot x)  \Sigma_x \right) d^4x\,,
\end{align}
where $\Sigma_x$ is the volume measure on $\mathcal{O}_x$ obtained from $\Sigma_x = \mathbf{i}_{\partial_0}\mathbf{i}_{\partial_1}\mathbf{i}_{\partial_2}\mathbf{i}_{\partial_3}\Sigma = \frac{\det g^L}{L^2}\mathbf{i}_{\mathbb{C}}(d^4\dot x)$. The extension of the integration from $V_x$ to $\mathcal{O}_x$ is always possible by the assumption of compact support of $\phi_x$.

Before we investigate the gravitational dynamics sourced by the kinetic gas in Section \ref{sec:GasGravity}, we now deduce the energy-momentum of the gas on the tangent bundle from manifold induced coordinate invariance of the action $S_{\text{gas}}$.

\subsection{General covariance and energy-momentum conservation}\label{ssec:EMDT}
All geometric objects in Finsler geometry have a distinguished behaviour under coordinate changes of the base manifold. That means even though the objects introduced in Section \ref{ssec:geom} are objects on the tangent bundle, their components in the adapted basis transform under coordinate changes of $TM$ induced by coordinate changes of $M$ just as if they were objects on $M$.

More precisely, consider a local coordinate change $x^a \mapsto \tilde x^a(x)$ on $M$. Such a coordinate change naturally induces a local coordinate change on $TM$ by $(x^a,\dot x^a)\mapsto(\tilde x^a, \dot {\tilde{x}}^a) = (\tilde x^a(x), \frac{\partial \tilde x^a}{\partial x^b}(x)\dot x^b)$. The basis change of the adapted horizontal and vertical bases are
\begin{align}
	\tilde \delta_a = \tilde \partial_a x^b \delta_b,\ \dot{\tilde{\partial}}_a = \tilde \partial_a x^b \dot{\partial}_b,\quad d\tilde x^a = \partial_b \tilde x^a dx^b,\ \delta \tilde{\dot x}^a = \partial_b \tilde x^a\delta \dot x^b\,.
\end{align}
Since all Finsler geometric objects are naturally expressed in the adapted basis, their components transform under manifold induced coordinate changes precisely the same way. All tensor fields on $TM$ with this property, are called $d$-, or distinguished, tensor fields, see \cite{Bucataru,Hohmann:2018hac} for details on d-tensors.

Since the Lagrangian $7$-form of the kinetic gas $\lambda = m \phi \Sigma$ is constructed out of a scalar and Finsler geometric objects, it is a $d$-$7$-form, and hence invariant under manifold induced coordinate changes. Studying infinitesimal manifold induced coordinate changes gives rise to conserved currents and an energy-momentum tensor of the gas on $TM$, which we will call \emph{energy-momentum distribution tensor $\Theta$}. It will turn out that this new tangent-bundle notion of energy-momentum distribution tensor of a kinetic gas is related to the usual definition of its energy-momentum tensor on the base manifold by averaging.

Infinitesimal manifold induced coordinate changes, labelled with a small expansion parameter $\epsilon$, are generated by a vector field $\xi=\xi^a(x) \partial_a$
\begin{align}
	\tilde x^a(x) = x^a + \epsilon \xi^a\,.
\end{align}
The corresponding manifold induced coordinate change on $TM$ is given by the additional change
\begin{align}
	 \dot{\tilde x}^a(x,\dot x) = \dot x^a +\epsilon \dot x^b \partial_b \xi^a\,.
\end{align}
Hence the total infinitesimal coordinate change on $TM$ is induced by the complete lift $\xi^C = \xi^a(x) \partial_a + \dot x^b \partial_b\xi^a \dot \partial_a$ of the vector field $\xi$. In the adapted basis $\xi^C = \xi^a \delta_a + \nabla\xi^a \dot{\partial}_a$ holds.

The Lagrangian $7$-form depends on the coordinates on $TM$ and on the field variable $L$. Hence its variation can be written as
\begin{align}
	\delta_{\xi} \lambda (x,\dot x, L,\dot{\partial}\dot{\partial}L) = \frac{d}{d\epsilon}\lambda (x + \epsilon \delta x,\dot x + \epsilon \delta \dot x, L + \epsilon \delta L, \dot{\partial}\dot{\partial}L + \epsilon \dot{\partial}\dot{\partial}\delta L)|_{\epsilon=0}\,,
\end{align}
where $\delta x^a = \xi^a,\ \delta \dot x^a = \dot x^b \partial_b\xi^a$ and $\delta L = -\xi^C(L)$. Evaluating the above derivative carefully, as it is done in Appendix~\ref{app:coordvari}, yields the expression
\begin{align}\label{eq:diffinvL}
	\delta_{\xi} \lambda = m\left(\mathfrak{L}_{\xi^C}(\phi\Sigma) + \delta_L (\phi \Sigma) \delta L\right)\,.
\end{align}
Using Cartan's magic formula $\mathfrak{L}_X = \mathbf{i}_{\xi^C}d + d\mathbf{i}_{\xi^C}$ in the first terms in \eqref{eq:diffinvL} reveals the equality
\begin{align}
	\mathfrak{L}_{\xi^C}(\phi\Sigma)
	&= \mathbf{i}_{\xi^C}d(\phi\Sigma)  + d\mathbf{i}_{\xi^C}(\phi\Sigma) = d\mathbf{i}_{\xi^C}(\phi\Sigma)\,,
\end{align}
where the term $\mathbf{i}_{\xi^C}d(\phi\Sigma)$ vanishes since \(\Sigma\) is a form of maximal rank on $\mathcal{O}$ and thus $d(\phi\Sigma)=0$.

The second term in \eqref{eq:diffinvL} gives
\begin{align}
	\delta_L(\phi \Sigma) \delta L
	&= (\delta_L(\phi \Omega) \wedge \omega + \phi \Omega \wedge \delta_L \omega) \delta L\nonumber\\
	&= (\phi \Omega \wedge \delta_L \omega)  \delta L\nonumber\\
	&= \frac{1}{2 L}\phi\ \delta L \ \Sigma \label{eq:VarL0}\\
	&= -\frac{1}{2 L}\phi\ \xi^C(L)\ \Sigma = -\frac{1}{2 L}\phi\ \dot{\partial}_a L \nabla\xi^a\ \Sigma\,.\label{eq:VarL}
\end{align}
In the first line we used $\Sigma = \Omega \wedge \omega$ and in the second line that $\delta_L(\phi\Omega)=0$, see \eqref{eq:dLN=0}. To reach the third line, the derivation of $\Omega \wedge \delta_L \omega = \frac{1}{2L}\phi \Sigma$ as it is displayed in Appendix \ref{app:Lvari} was applied. Finally, for the variation with respect to the manifold induced coordinates, we have that $\delta L = -\xi^C(L) = \nabla \xi^a \dot{\partial}_aL$, since $\delta_a L = 0$. Inserting $\dot{\partial}_aL = 2 g^L_{ab}(x,\dot x)\dot x^b = 2 \dot x_a$, the total change of the Lagrange $7$-form of the kinetic gas is given by
\begin{align}
	\delta_{\xi} \lambda = m \left( -\phi\frac{\dot x_a}{L} \nabla \xi^a\ \Sigma  +  d\mathbf{i}_{\xi^C} (\phi \Sigma)\right)\,,
\end{align}
and so the variation of the action is
\begin{align}\label{eq:vargasact}
	\delta_{\xi}S_{\text{gas}} = - m \int_V \phi\frac{\dot x_a}{L} \nabla \xi^a\ \Sigma  + m \int_{\partial V}\mathbf{i}_{\xi^C}(\phi \Sigma)\,.
\end{align}
From equation \eqref{eq:VarL0} we see that the integrand of the volume integral in \eqref{eq:vargasact} is the Euler-Lagrange part of the variation of the action with respect to the Finsler Lagrange function $L$, which will define the source term of the gravitational dynamics in the next section. The boundary term, where $\partial V = (\bigcup_{x\in D}\partial V_x) \cup (\bigcup_{x\in \partial D} V_x)$, compare \eqref{eq:actsplit}, can thus be be interpreted as Noether current\footnote{Actually, the current $J^\xi$ is only a partial Noether current since the contribution of the gravitational Lagrangian, giving dynamics to the Finsler Lagrangian $L$, needs to be added.} $J^\xi =  \mathbf{i}_{\xi^C} (\phi\Sigma)$ associated to invariance of the action \eqref{eq:gasact} under the $1$-parameter group of $\xi^C$ \cite{Voicu:EM}.

To identify the energy-momentum tensor of the kinetic gas on the tangent bundle, we use the product rule in the first term of the variation of the action and obtain
\begin{align}
	\delta_{\xi}S_{\text{gas}} = - m \int_V \nabla \left( \frac{\phi \dot x_a}{L} \xi^a\right) \Sigma + m \int_V \nabla_{\delta_b} \left( \frac{\phi \dot x_a \dot x^b}{L} \right) \xi^a \Sigma + m \int_{\partial V} \mathbf{i}_{\xi^C}(\phi \Sigma)\,,
\end{align}
where $\nabla_{\delta_b}$ and $\nabla$ are the Chern-Rund and the dynamical covariant derivative introduced in \eqref{eq:CRCD} and \eqref{eq:dynCD}.

The first term is again a boundary term, which can be combined with the last term setting $f:=\frac{\phi \dot x_a \xi^a}{L}$
\begin{align}\label{eq:dSgas}
\delta_{\xi}S_{\text{gas}} = m \int_V \nabla_{\delta_b} \left( \frac{\phi \dot x_a \dot x^b}{L} \right) \xi^a \Sigma + m \int_{\partial V} (\mathbf{i}_{\xi^C}(\phi \Sigma) - \mathbf{i}_{f\mathbf{r}}\Sigma)\,.
\end{align}
Using the methods from \cite{GotayMarsden,Voicu:EM}, we find that the middle term singles out the desired \emph{energy-momentum distribution tensor}, which is defined as
\begin{align}\label{eq:EMDT}
	\Theta^a{}_b(x,\dot x) = m \frac{\phi \dot x^a \dot x_b}{L}\,.
\end{align}

Having identified the canonical energy-momentum distribution tensor of a kinetic gas on the tangent bundle let us us make four important observations:
\begin{itemize}
	\item The average of $\Theta^a{}_b(x,\dot x)$ over all observer directions $\mathcal{O}_x$ at a point $x\in M$ gives the components $T^a{}_b(x)$ of the energy-momentum tensor density of the gas on $M$
	\begin{align}\label{eq:conlawa}
		\int_{\mathcal{O}_x} \Theta^a{}_b(x,\dot x) \Sigma_{x} = \int_{\mathcal{O}_x} m \frac{\phi \dot x^a \dot x_b}{L}\Sigma_{x} = T^a{}_b(x)\,.
	\end{align}
	The last equality makes sense by the assumption that $\phi_x = \phi(x,\cdot)$ is compactly supported. In case the Finsler geometry is is pseudo-Riemannian, i.e. $g^L_{ab}(x,\dot x) = g_{ab}(x)$, the lower index velocity $\dot x_b$ can be written as $g_{bc}(x)\dot x^c$, the metric components can be pulled out of the integral and $T^a{}_b(x)$ can be written as $T^{ac}(x)g_{bc}(x)\sqrt{\det (g_{mn}(x))}$ where $T^{ac}(x)$ is identical to the energy-momentum tensor of a kinetic gas in the literature, obtained as second moment of the 1PDF with respect to the four-velocities \cite{Ehlers2011,Sarbach:2013fya}. In the general Finsler setting, generically, there exists no metric tensor on the spacetime manifold $M$, so it is not possible to raise or lower an index on $T^a{}_b(x)$ and thus one needs to work with the canonically defined $(1,1)$-tensor density field.

	\item Rewriting  \eqref{eq:dSgas} as an iterated integral over a compact domain $D \subset M$, and a compact domain $V_x \subset \mathcal{O}_x$ which contains $\textrm{supp}(\phi_x)$, see \eqref{eq:actsplit}, yields
	\begin{align}
	0
	&= \int_D \left( \int_{V_x} \nabla_{\delta_b} \Theta^b{}_a(x,\dot x) \xi^a \Sigma_x\right) d^4x +  m \int_{(\bigcup_{x\in D}\partial V_x) \cup (\bigcup_{x\in \partial D} V_x)} (\mathbf{i}_{\xi^C}(\phi \Sigma) - \mathbf{i}_{f\mathbf{r}}\Sigma)\nonumber\\
	&= \int_D \left( \int_{\mathcal{O}_x} \nabla_{\delta_b} \Theta^b{}_a(x,\dot x) \xi^a \Sigma_x\right) d^4x +  m \int_{(\bigcup_{x\in D}\partial \mathcal{O}_x) \cup (\bigcup_{x\in \partial D} \mathcal{O}_x)} (\mathbf{i}_{\xi^C}(\phi \Sigma) - \mathbf{i}_{f\mathbf{r}}\Sigma)\,.
	\end{align}
	The extension of the integration from $V_x$ to all of $\mathcal{O}_x$ in the second line is viable due to the compact support of~$\phi_x$. Since $\mathcal{O}_x$ has no boundary and the invariance $\delta_\xi S_{\text{gas}} = 0$ holds in particular for compactly supported vector fields $\xi$ whose support is contained in $D$, i.e. $\xi|_{\partial D} = 0$, we find that the boundary term vanishes and hence at each point $x\in M$, the averaged conservation law
	\begin{align}\label{eq:conlaw1}
		\int_{\mathcal{O}_x} \nabla_{\delta_b} \Theta^b{}_a \Sigma_x = 0\,.
	\end{align}

	\item For collisionless gas a non-averaged conservation law holds
	\begin{align}\label{eq:conlaw2}
		\nabla_{\delta_b} \Theta^b{}_a = \frac{\dot x^b \dot x_a}{L} \nabla_{\delta_b} \phi = \frac{\dot x^b \dot x_a}{L} \delta_b \phi =\frac{\dot x_a}{L} \mathbf{r}(\phi) = 0\,,
	\end{align}
	since $\nabla_{\delta_b} \dot x^a = 0$, $\nabla_{\delta_b} L = 0$, $\nabla_{\delta_b} g^L_{ac} = 0$, by construction of the Chern-Rund covariant derivative, see \eqref{eq:CRCD} and Appendix \ref{app:CRConn}, and the fact that the 1PDF of a collisonless gas satisfies the Liouville equation $\mathbf{r}(\phi) = 0$, see \eqref{eq:lville}. Thus the Liouville equation can be interpreted as covariant energy-momentum distribution tensor conservation equation.

	\item Using the averaged conservation law \eqref{eq:conlaw1} in \eqref{eq:dSgas}, using $f=\frac{\phi \dot x_a \xi^a}{L}$, we find for arbitrary $\xi$, only the horizontal part of $\xi^C$ contributes to the integral,
	\begin{align}\label{eq:JT}
		\int_{(\bigcup_{x\in D}\partial \mathcal{O}_x) \cup (\bigcup_{x\in \partial D} \mathcal{O}_x)} ( \mathbf{i}_{\xi^C}(\phi \Sigma) - \Theta^b{}_a(x,\dot x)\xi^a\mathbf{i}_{\delta_b}\Sigma)
		=\int_{\bigcup_{x\in \partial D} \mathcal{O}_x} ( \xi^b\mathbf{i}_{\delta_b}(\phi \Sigma) - \Theta^b{}_a(x,\dot x)\xi^a\mathbf{i}_{\delta_b}\Sigma)  = 0\,.
	\end{align}
	The above equality, proven in Appendix~\ref{app:3}, relates the energy-momentum distribution tensor to the previously identified quasi Noether current. This demonstrates that indeed $\Theta^a{}_b(x,\dot x)\xi^b$ represents the \emph{corrected Noether current}, see \cite{GotayMarsden}, due to general covariance of the Lagrangian, that is, it provides the correct energy-momentum of the system \cite{GotayMarsden,Voicu:EM}.
\end{itemize}

In summary our findings clearly demonstrate that the description of kinetic gases on the base manifold is just a velocity averaged description. The finer, more precise non-averaged description of the kinetic gas can be formulated on the tangent bundle.

In the next section we couple the tangent bundle description of the kinetic gas directly to the gravity.

\section{Coupling the kinetic gas to gravity}\label{sec:GasGravity}
The idea that Finsler geometry is a good candidate to generalise general relativity is around for a long time \cite{Asanov,Rutz,Li2014,Pfeifer,Hohmann:2018rpp,Minguzzi:2014fxa}. One highly debated question is how to couple matter fields consistently to a Finsler geometric theory of gravity, which naturally lives on the tangent bundle of spacetime instead of on spacetime itself. Usually, this feature is seen as a complication to construct a consistent complete theory.

After our discussion of the kinetic gas in the language of Finsler geometry, we will turn this complication into an advantage. The best motivated Finsler gravity action was formulated on the projectivised tangent bundle. In particular, taking care of all mathematical details, which are discussed in the recent article \cite{Hohmann:2018rpp}, it was shown that it can be equivalently understood as a theory on the unit tangent bundle, which contains the observer space $\mathcal{O}$. Therefore, Finsler gravity offers a direct coupling of gravity to the Finslerian description of a gravitating relativistic kinetic gas.

For this purpose, we propose the following action:
\begin{align}
S = \frac{1}{2\kappa^2} S_{\text{grav}} + S_{\text{gas}} = \frac{1}{2\kappa^2} \int_V R_0 \Sigma + \int_V \phi \Sigma\,,
\end{align}
where $\kappa$ is the Finslerian gravitational coupling constant.

To obtain the gravitational field equations we perform variation of the action with respect to $L$. For the gravitational part we found in \cite{Pfeifer,Hohmann:2018rpp}
\begin{align}
	\delta_L S_{\text{grav}} = \int_V \bigg[\frac{1}{2}g^{Lab}\dot{\partial}_a\dot{\partial}_b(L R_0) - 3 R_0 - g^{Lab}(\nabla_{\delta_a}P_{b} - P_aP_b + \dot{\partial}_a(\nabla P_b)) \bigg]\frac{\delta L}{L}\Sigma\,,
\end{align}
where $P_a = P^{b}{}_{ba}$ are the components of the trace of the Landsberg tensor, which was defined in \eqref{eq:Lands}.

Variation of the our newly constructed kinetic gas action yields
\begin{align}\label{eq:dlSg}
\delta_L S_{\text{gas}} = \int_V \delta_L(\phi \Omega \wedge \omega) = \int_V \delta_L(\phi \Omega) \wedge \omega + \int_V \phi \Omega \wedge \delta_L \omega\,.
\end{align}
The first term is nothing but the variation of the number counting integral and thus vanishes, see \eqref{eq:dLN=0},
\begin{align}
	\int_V \delta_L(\phi \Omega) \wedge \omega = 0\,.
\end{align}
In Appendix \ref{app:Lvari} we derive that
\begin{align}
	\Omega \wedge \delta_L\omega = \frac{1}{2}\frac{\delta L}{L} \Sigma\,,
\end{align}
and thus obtain from \eqref{eq:dlSg}
\begin{align}
	\delta_L S_{\text{gas}} = \int_V\phi \frac{1}{2}\frac{\delta L}{L} \Sigma\,.
\end{align}

Eventually, the Finsler gravity equations sourced by a kinetic gas on the observer space $\mathcal{O}$ are
\begin{align}\label{eq:fgravgas}
	\frac{1}{2}g^{Lab}\dot{\partial}_a\dot{\partial}_b(R_0 L) - 3 R_0 - g^{Lab}(\nabla_{\delta_a}P_{b} - P_aP_b + \dot{\partial}_a(\nabla P_b)) = -\kappa^2\phi\,.
\end{align}
This equation determines the geometry of spacetime, i.e. the gravitational field, directly from the 1PDF of a kinetic gas. It takes into account the influence of the in general non-trivial velocity distribution over the different gas particles. The first two terms appearing can be understood analogue to the appearance of the Ricci tensor and Ricci scalar in the Einstein tensor, in the sense that these terms arise from the traces of the geodesic deviation operator of the spacetime geometry. The term involving the trace of the Landsberg tensor, which measures the change of the Cartan tensor along Finsler geodesics, is purely Finslerian. The precise physical meaning of these terms is an ongoing investigation.

In contrast to the Einstein equations, our new equation is not a tensorial equation for a metric on spacetime, but a scalar equation for a Finsler Lagrangian on the tangent bundle of spacetime, as it must be when one considers a \emph{scalar function} on the tangent bundle as fundamental gravitational field variable. A natural question appearing is, how and if this equation is related to the Einstein equations. For the vacuum equation we demonstrated in \cite{Hohmann:2018rpp,Pfeifer} that they are equivalent to the Einstein vacuum equations if a Finsler Lagrangian of the type $L(x,\dot x)=g_{ab}(x)\dot x^a \dot x^b$ is considered.

To discuss this question for the matter coupled equation we employ the relation between the fluid energy-momentum tensor and the 1PDF \eqref{eq:conlawa}. We apply the same integration operator to both sides of our field equation \eqref{eq:fgravgas},
\begin{align}
\int_{\mathcal{O}_x}\frac{\dot x^c\dot x_d}{L}\left(\frac{1}{2}g^{Lab}\dot{\partial}_a\dot{\partial}_bR - 3 \frac{R}{L} - g^{Lab}(\nabla_{\delta_a}P_{b} - P_aP_b + \dot{\partial}_a(\nabla P_b))\right)\Sigma_x
= - \int_{\mathcal{O}_x} \kappa^2 \frac{\dot x^c\dot x_d}{L}\phi \Sigma_x = - \frac{\kappa^2}{m} T^c{}_d(x)\,.
\end{align}
This equation is a tensorial equation of the type $\mathfrak{G}^a{}_b = - \frac{\kappa^2}{m} T^a{}_b(x)$ on the spacetime manifold. An open question, which we are currently investigating, is, if, how and under which conditions the tensor field $\mathfrak{G}^a{}_b$, resulting from velocity averaging, is related to the Einstein tensor.

We like to stress that the above equation already looks close to the Einstein-Vlasov system, which is one way how the gravitational field of a kinetic-gas is determined in the literature \cite{Andreasson:2011ng}. Denote with $g_{ab}=g_{ab}(x)$ the components of a pseudo-Riemannian spacetime metric and with $r_{ab}$ and $r$ the components of the Ricci tensor and Ricci scalar of its Levi-Civita connection. Then the Einstein-Vlasov equation can be written as\footnote{Often, the Einstein-Vlasov system is considered on the cotangent bundle of spacetime, i.e.\ using momenta instead of velocities. As long as a Legendre transform between the Lagrangian and Hamiltonian formulation of the underlying point particle mechanics is available both pictures are equivalent.}
\begin{align}
	r^a{}_b - \frac{1}{2}\delta^a{}_b r = \frac{8\pi G}{c^4} \frac{1}{\sqrt{|\det g|}} \int_{\mathcal{O}_x}  \frac{\dot x^a\dot x_b}{g_{ij}(x)\dot x^i \dot x^j}\phi\ \Sigma_x\,,
\end{align}
where here $\Sigma_x$, as defined below equation \eqref{eq:actsplit}, is nothing but $\sqrt{\det g}$ times the canonical volume form on $\mathcal{O}_x$ induced by the pullback of the spacetime metric.

Comparing the Einstein-Vlasov system and our new approach to the determination of the gravitational field of a $P$-particle system, we conjecture the following interpretation: The Einstein equations determine only the velocity averaged gravitational field of the $P$-particle system. Our Finslerian description of the gravitational field of kinetic gases determines the gravitational field distribution, without averaging over the velocity distribution of the particles, in the same way as the 1PDF describes the system more accurately than its velocity averaged fluid approximation.

\section{Conclusion}\label{sec:conc}

We used the language of Finsler geometry to describe a multiple particle system as a kinetic gas in terms of a 1PDF on the tangent bundle, which we directly coupled to a Finslerian extension of Einstein's theory of gravity. In contrast to the usual general relativistic coupling between the multi particle system and gravity, via its fluid energy-momentum tensor as source of the Einstein equations, our new coupling procedure takes the velocity distribution of the gas particles and its influence on the gravitational field into account. In contrast, on the right hand side of the Einstein equations, the information of the velocity distribution is averaged out.

Using the language of Finsler spacetime geometry, we obtained the main results of this article: our finding of the energy-momentum distribution tensor  \eqref{eq:EMDT} and its conservation laws for a kinetic gas with and without collisions as displayed in \eqref{eq:conlaw1} and \eqref{eq:conlaw2}, and the Finsler generalizations of the Einstein equations sourced by the 1PDF of the kinetic gas in \eqref{eq:fgravgas}.

With the construction of an action based formulation of the dynamics of a kinetic gas and its coupling to gravity, we complete the Finsler geometric description of kinetic gases and are now able to derive the full mutual kinetic gas-gravity interaction on the tangent bundle. Moreover, by adding this physical matter coupling to Finsler gravity, a most subtle point in the construction of Finslerian theories of gravity - the matter coupling - is solved.


Our work offers a plethora of possible applications. One may apply our extended description of gravity and kinetic gases to any physical system which is conventionally described by classical fluid dynamics. An obvious system to target with this description is of course cosmology, where a better model of fluids and gravity may provide potential new explanations for the observed accelerating phases, known as dark energy and inflation, via a modification of the effective Friedmann equations. Another possible application is to the dynamics of galaxies and the large scale structure of the Universe, where conventionally dark matter must be assumed in order to explain observations. Further, compact objects such as neutron stars may provide a potential testbed for our theory, since the description of their constituting matter is crucial for their understanding.

The next step towards these applications is to identify physically well motivated 1PDFs, which may be for example given by a tangent bundle version of the relativistic Maxwell-Juettner Distribution \cite{Treumann:2015etb}, which is usually formulated on the cotangent bundle. Other fascinating possibilities are to find 1PDF realisations of quantum distribution functions such as Bose-Einstein or Fermi-Dirac distributions and couple them directly to gravity.

A further interesting line of investigation is to study the relation of our description of the relativistic kinetic gas to the framework of kinetic field theory, which was recently applied to cosmology \cite{Bartelmann:2019unp}.

We leave such investigations for future research, keeping the discussion presented in this article at the foundational level, having constructed the gravitational field equations for a gravitating kinetic gas.

\begin{acknowledgments}
MH and CP were supported by the Estonian Ministry for Education and Science through the Personal Research Funding Grant PRG356, as well as the European Regional Development Fund through the Center of Excellence TK133 ``The Dark Side of the Universe''. The authors would like to acknowledge networking support by the COST Actions CANTATA (CA15117) and QGMM (CA18108), supported by COST (European Cooperation in Science and Technology).
\end{acknowledgments}

\appendix
\section{Properties of the Chern-Rund covariant derivative}\label{app:CRConn}
For completeness, we give a proof of the basic properties of the Chern-Rund covariant derivative \eqref{eq:CRprop}.

The covariant constancy of the $L$-metric can easily be seen from the definition of the connection coefficients $\Gamma^c{}_{ab} = \frac{1}{2}g^{Lcq}(\delta_a g^L_{bq}+\delta_b g^L_{aq}-\delta_q g^L_{ab})$
\begin{align}
\nabla_{\delta_a}g^L_{bc} = \delta_a g^L_{bc} - \Gamma^p{}_{ab}g^L_{bc} - \Gamma^p{}_{ac}g^L_{bp} = 0\,.
\end{align}

The covariant constancy of $\dot x^a$ requires some more lines.
\begin{align}
	\nabla_{\delta_a}\dot x^b = \delta_a\dot x^b + \Gamma^b{}_{ac}\dot x^c = (\partial_a - G^c{}_a\dot \partial_c)\dot x^b + \Gamma^b{}_{ac}\dot x^c  = - G^b{}_a + \Gamma^b{}_{ac}\dot x^c =0\,.
\end{align}
The last equality above can be seen from
\begin{align}
	\Gamma^b{}_{ac}\dot x^c
	&=  \frac{1}{2}g^{Lbq}(\delta_a g^L_{cq}+\delta_c g^L_{aq}-\delta_q g^L_{ac})\dot x^c \nonumber\\
	&= \frac{1}{2}g^{Lbq}(\partial_a g^L_{cq}+\partial_c g^L_{aq}-\partial_q g^L_{ac})\dot x^c - \frac{1}{2}g^{Lbq}G^p{}_c \dot x^c \dot{\partial}_p g^L_{qa} \nonumber\\
	&= \frac{1}{4}g^{Lbq}(\partial_a \dot{\partial}_q L + \dot x^c\partial_c g^L_{aq}-\partial_q \dot \partial_a L) - g^{Lbq} G^p \dot{\partial}_a g^L_{qp} \nonumber\\
	&= \frac{1}{4}g^{Lbq}(\partial_a \dot{\partial}_q L + \dot x^c\partial_c g^L_{aq}-\partial_q \dot \partial_a L) - g^{Lbq} \dot{\partial}_a(G^p  g^L_{qp}) + G^b{}_a \nonumber\\
	&= G^b{}_a\,,
\end{align}
where we used the definition of the $\delta_a$ operator, several times Euler's Theorem for homogeneous functions, the total symmetry in all indices of $\dot{\partial}_a g^L_{qp}$ and the definition of $G^a$ and $G^a{}_b$, see \eqref{def:Ga}.

Finally the covariant constancy of $L$ itself is a simple consequence of the above equalities
\begin{align}
	\nabla_{\delta_a} L = \nabla_{\delta_a} (g^L_{bc}\dot x ^b \dot x^c) = 0\,.
\end{align}

\section{Variation of the kinetic gas action under manifold induced coordinate changes}\label{app:coordvari}
When we studied the variation of the action of the kinetic gas under manifold induced coordinate changes we claimed equation \eqref{eq:diffinvL}
\begin{align}
		\delta_{\xi} \lambda = m(\mathfrak{L}_{\xi^C}(\phi\Sigma) + \delta_L (\phi \Sigma) \xi^C(L))\,.
\end{align}
which we will prove now.

We start from the definition of the variation
\begin{align}
	\delta_{\xi} \lambda (x,\dot x, L, \dot{\partial}\dot{\partial}L) = \frac{d}{d\epsilon}\lambda (x + \epsilon \delta x,\dot x + \epsilon \delta \dot x, L + \epsilon \delta L, \dot{\partial}\dot{\partial}L + \epsilon \dot{\partial}\dot{\partial}\delta L)|_{\epsilon=0}\,,
\end{align}
where $\delta x^a = \xi^a,\ \delta \dot x^a = \dot x^b \partial_b\xi^a$ and $\delta L = -\xi^C(L)$.

First, we notice that the Lagrangian $7$-form in consideration, \eqref{eq:L7form}, can be split into a coordinate volume form and a Lagrange scalar density $\mathcal{L}$
\begin{align}
	\lambda = \mathcal{L}(x,\dot x, L, \dot{\partial}\dot{\partial}L) \mathbf{i}_{\mathbf{C}}(d^4x\wedge d^4\dot x)\,.
\end{align}

The change of the coordinate volume form can be calculated by means of the identities
\begin{align}
	d(x^a + \epsilon \delta x^a) &= d x^a + \epsilon \partial_b \delta x^a dx^b = d x^a + \epsilon \partial_b \xi^a dx^b\,,\\
	d(\dot x^a + \epsilon \delta \dot x^a) &= d \dot x^a + \epsilon (\partial_b \delta \dot x^a dx^b + \dot \partial_b \delta \dot x^a d\dot x^b) = d \dot x^a + \epsilon ( \dot x^c \partial_b \partial_c\xi^a dx^b + \partial_b\xi^a d\dot x^b)\,.
\end{align}
For the wedge product of these one forms we thus get to first order in $\epsilon$
\begin{align}
	\mathbf{i}_\mathbb{C}(d^4(x+\epsilon \delta x)\wedge d^4(\dot x+\epsilon \delta \dot x))
	&= \det
    \left(
	\begin{array}{c|c}
	\partial_b ( x^a+\epsilon \delta x^a) & \dot \partial_b ( x^a+\epsilon \delta x^a) \\ \hline
	\partial_b (\dot x^a+\epsilon \delta \dot x^a) & \dot \partial_b (\dot x^a+\epsilon \delta \dot x^a)
	\end{array}
	\right)\mathbf{i}_\mathbb{C}(d^4x\wedge d^4\dot x)\nonumber\\
	&= \det
	\left(
	\begin{array}{c|c}
	\delta^a_b + \epsilon \partial_b \xi^a & 0\\ \hline
	\epsilon \dot x^c \partial_b \partial_c \xi^a & \delta^a_b + \epsilon \dot \partial_b (\dot x^c \partial_c\xi^a)
	\end{array}
	\right)(d^4x\wedge d^4\dot x)\nonumber\\
	&= \left(1 + \epsilon (\partial_a \xi ^a + \dot \partial_b (\dot x^c \partial_c \xi^b) )\right)\mathbf{i}_\mathbb{C}(d^4x\wedge d^4\dot x)\,.
\end{align}

Second, direct Taylor expansion to first order in $\epsilon$ yields the change of the scalar density $\mathcal{L}$ under the coordinate change to be
\begin{align}
	\mathcal{L}(x + \epsilon \delta x,\dot x + \epsilon \delta \dot x, L + \epsilon \delta L, \dot{\partial}\dot{\partial}L + \epsilon \dot{\partial}\dot{\partial}\delta L)
	= \mathcal{L}(x,\dot x, L, \dot{\partial}\dot{\partial}L) + \epsilon (\xi^C(\mathcal{L}) - \delta_L \mathcal{L}\ \xi^C(L))|_{\mathcal{L}=\mathcal{L}(x,\dot x, L, \dot{\partial}\dot{\partial}L)}\,.
\end{align}

We did not perform the expansion of the derivatives with respect to $\epsilon$ of the terms involving the change of $L$ explicitly but combined them into the term $\delta_L \mathcal{L}$, since this variation can be done most simply after we recombine the terms and get
\begin{align}
	\delta_{\xi}\lambda
	&= \left( \mathcal{L}\ (\partial_a \xi ^a + \dot \partial_b (\dot x^c \partial_c \xi^b)) + \xi^C(\mathcal{L}) \right)\mathbf{i}_\mathbb{C}(d^4x\wedge d^4\dot x) - \delta_L\mathcal{L}\ \xi^C(L)\mathbf{i}_\mathbb{C}(d^4x\wedge d^4\dot x)\nonumber\\
	&= \mathfrak{L}_{\xi^C}\lambda -  \xi^C(L) \delta_L\lambda
	= d\mathbf{i}_{\xi^C}\lambda - \xi^C(L) \delta_L\lambda \,.
\end{align}
Inserting $\lambda = m\phi\Sigma$ proves the desired variation formula \eqref{eq:diffinvL}. The variation $\delta_L\lambda = m \delta_L(\phi\Sigma)$, which comes from the change of $L$, is discussed during the derivation of the gravitational field equation coupled to the gas in Section \ref{sec:GasGravity} and Appendix \ref{app:Lvari}.

\section{Proof of the variation of the kinetic gas action with respect to L}\label{app:Lvari}
During the variation of the action of the kinetic gas with respect to $L$, \eqref{eq:VarL} and \eqref{eq:dlSg}, we evaluated
\begin{align}\label{eq:vargasactap}
	\phi \Omega \wedge \delta_L \omega = \phi \frac{1}{2}\frac{\delta L}{L}\Sigma\,.
\end{align}

To prove this equation, we first observe that on $\mathcal{O}$
\begin{align}
\delta_L \omega = \delta_L \dot{\partial}_a F dx^a = \dot{\partial}_a \delta_L  F dx^a = \dot{\partial}_a\left(\frac{1}{2}\frac{\delta L}{\sqrt{L}}\right)dx^a\,.
\end{align}
Furthermore, for functions $f$ on $TM$, which are $h$-homogeneous with respect to $\dot x$ and at least $\mathcal{C}^1$, the following holds:
\begin{align}
	\Omega \wedge \dot{\partial}_sf dx^s
	&= \mathbf{i}_\mathbf{r}\Sigma \wedge \dot{\partial}_s f dx^s
	= \mathbf{i}_\mathbf{r}(\Sigma \wedge \dot{\partial}_s f dx^s) + \Sigma \wedge \dot{\partial}_s f \mathbf{i}_\mathbf{r}dx^s\nonumber\\
	&= \dot x^s \dot{\partial}_s f \Sigma = h f \Sigma\,,
\end{align}
where we used the product rule for the interior product, the relation $\Sigma \wedge dx^s = 0$ satisfied by the $7$-form $\Sigma$ and Euler's theorem for homogeneous functions. Inserting $f = \frac{1}{2}\frac{\delta L}{\sqrt{L}}$ and $h=1$ we find
\begin{align}
	\Omega \wedge \dot{\partial}_sf dx^s = \Omega \wedge \delta_L \omega
	&=\frac{1}{2} \frac{\delta L}{L} \Sigma\,.
\end{align}

\section{Integration of the vertical interior product}\label{app:3}
When we found that the energy-momentum distribution tensor is related to the Noether current in equation \eqref{eq:JT}, we used that
\begin{align}
\int_{(\bigcup_{x\in D}\partial \mathcal{O}_x) \cup (\bigcup_{x\in \partial D} \mathcal{O}_x)} (\nabla\xi^a \mathbf{i}_{ \dot \partial_a}(\phi \Sigma))=0
\end{align}
to equate
\begin{align}\label{eq:JTapp}
\int_{(\bigcup_{x\in D}\partial \mathcal{O}_x) \cup (\bigcup_{x\in \partial D} \mathcal{O}_x)} ( \mathbf{i}_{\xi^C}(\phi \Sigma) - \Theta^b{}_a(x,\dot x)\xi^a\mathbf{i}_{\delta_b}\Sigma)
=\int_{\bigcup_{x\in \partial D} \mathcal{O}_x} ( \xi^b\mathbf{i}_{\delta_b}(\phi \Sigma) - \Theta^b{}_a(x,\dot x)\xi^a\mathbf{i}_{\delta_b}\Sigma)  = 0\,,
\end{align}
since in the first term $\mathbf{i}_{\xi^C}(\phi \Sigma) = \xi^b \mathbf{i}_{\delta_b}(\phi \Sigma) + \nabla\xi^a\mathbf{i}_{ \dot \partial_a}(\phi \Sigma)$.

The best way to see this is to rewrite the integral as a volume integral and use the splitting into an iterated integral over a domain $D\subset M$ and the observer space fibre $\mathcal{O}_x$
\begin{align}
	\int_{(\bigcup_{x\in D}\partial \mathcal{O}_x) \cup (\bigcup_{x\in \partial D} \mathcal{O}_x)} \nabla\xi^a\mathbf{i}_{ \dot \partial_a}(\phi \Sigma)
	= \int_V d \mathbf{i}_{ \nabla\xi^a\dot \partial_a}(\phi \Sigma)
	= \int_V \textrm{div}(\nabla\xi^a\dot \partial_a)\Sigma
	= \int_D \left( \int_{\mathcal{O}_x} \textrm{div}(\nabla\xi^a\dot \partial_a)\Sigma_x\right)d^4x\,
\end{align}
The vector field $\nabla\xi^a\dot \partial_a$ is purely vertical and hence the inner integral becomes an integral over the boundary of $\mathcal{O}_x$, which is empty. Hence the whole integral vanishes
\begin{align}
	\int_{\mathcal{O}_x} \textrm{div}(\nabla\xi^a\dot \partial_a)\Sigma_x = \int_{\partial \mathcal{O}_x} \mathbf{i}_{\nabla\xi^a\dot \partial_a} \Sigma_x = 0\,.
\end{align}

\bibliographystyle{utphys}
\bibliography{KGG}

\end{document}